\documentclass[floatfix,showpacs,aps,nofootinbib,onecolumn]{revtex4}

\usepackage[T1]{fontenc}
\usepackage[english]{babel}
\usepackage[utf8]{inputenc}
\usepackage{amsmath,amsfonts,amssymb,mathrsfs}
\usepackage{xcolor}
\usepackage{yfonts}

\begin{document}

\title{A pilot study on canonical gravity with mass dimension one fermions}

\author{R. de C. Lima$^{1}$}\email{rodrigo.lima@ifnmg.edu.br}
\author{T. M. Guimar\~aes$^{2}$}\email{thiago.moreira@ifpr.edu.br}
\author{S. H. Pereira$^{3}$}\email{s.pereira@unesp.br}

\affiliation{$^{1}$Instituto Federal de Educa\c{c}\~ao, Ci\^encia e Tecnologia do Norte de Minas Gerais (IFNMG), Janu\'aria, MG, 39480-000, Brazil.}
\affiliation{$^{2}$Instituto Federal de Educa\c{c}\~ao, Ci\^encia e Tecnologia do Paran\'a (IFPR), Ivaipor\~a, PR, 86870-000, Brazil.}
\affiliation{$^{3}$Departamento de F\'isica, Faculdade de Engenharia e Ci\^encias de Guaratinguet\'a, Universidade Estadual Paulista (UNESP), Guaratinguet\'a, SP, 12516-410, Brazil.}


\begin{abstract}\vspace{0.15cm}
\vspace{0.3cm}
\begin{center}
\textbf{Abstract} 
\end{center}
The  mass dimension one (MDO) fermionic field is built on a complete set of dual-helicity eigenspinors of the charge conjugation operator, which obeys the statistic of Fermi-Dirac. These spinors are a potential candidates for the description of dark matter. The redefinition of the dual structure of this object holds a local adjacent theory and completely satisfies the Lorentz invariance. In this work, we investigate important aspects of the interaction of this fermion with gravity in the light of a canonical formulation in ADM formalism. We construct an action via tetrad fields using a manifold on a family of space-like surfaces $\Sigma_{t}$ that carries the MDO matter field, and additionally we propose a condition for the action to have a term associated with dark energy. We found the Hamiltonian and diffeomorphism constraints at the classical level for gravitational dynamics with the immersion of this material content in space-time, which leads us to the interpretation of the contribution of dark matter energy density, parallel to the lapse function of foliation and its directional flux of energy density in the hypersurface of manifold.
\vspace{0.45cm}
\vspace{0.8cm} 
\end{abstract}


\maketitle
\section{Introduction}\label{S1}
\indent
\indent Proposed in its first version in $2005$, the fermionic field of spin-$1/2$ with mass dimension one, abbreviated to MDO, is built on a complete set of eigenspinors of the charge conjugation operator $\mathcal{C}$, the so called \textit{Elko}\footnote{Elko is a german acronym for \textit{Eigenspinoren des
Ladungskonjugationsoperators}} spinors \cite{Ahluwalia:2004ab}. The construction of the field is characterized by the presence of spinors belonging to the $(0,1/2) \oplus (1/2,0)$ family of the Weyl representation space. In its original formulation, these fields were quantum objects based on a representation of the subgroups $HOM (2)$ and $SIM (2)$ of the Lorentz group, whose algebra is based on Very Special Relativity (VSR) \cite{Cohen:2006ky}. A few years ago, a modification of the dual of the field - taking advantage of the fact that only bilinears are observable in nature - endowed the field with complete Lorentz  (Poincaré) symmetry \cite{ Ahluwalia:2016rwl}. Thus, the old non-local field turns into a local one after a redefinition of phase factors in the right and left-hand components of the field. These recent developments put the theory on solid grounds and stimulated several works in broad areas of Physics, with numerous results explored in the book \cite{Ahluwalia:2019etz}, in Particle Phenomenology \cite{Duarte:2020svn} and Cosmology \cite{Pereira:2020ogo}, as brief examples. More recently, Ref. \cite{Ahluwalia:2022ttu} presents a detailed discussion of duals and adjoints of the field, and the fermionic self-interaction and interactions with a real scalar field at one-loop shows that the earlier problem of unitarity violation are absent. Also, a quantum field theoretic calculation establishes the Newtonian gravitational interaction for a mass dimension one dark matter candidate. The partition function and main thermodynamic properties were studied and a review on the localization of higher-dimensional ELKOs on flat and bent branes were discussed.

The crucial difference between Dirac and MDO spinors is related to parity. For Dirac spinors, the parity is intrinsic to the theory and, as a direct consequence, Dirac dynamics is achieved. For the MDO spinors this does not happen, as well analyzed in \cite{CoronadoVillalobos:2016lkj}. However, since its construction is relativistic, the Klein-Gordon dynamics is ensured, thus, its quantum field must inherit this dynamic. The MDO quantum field is constructed as an expansion in terms of the four different ELKO spinors, namely two self-conjugated and two anti self-conjugated. These spinors are eigenspinor of the charge-conjugation operator, what ensures they are electrically neutral. Furthermore, it has become a prominent theoretical dark matter candidate, in parts, due its canonical mass dimension, $\mathfrak{D} = 1$, similar to scalar particles, what forbidden interactions of this field with other standard model particles, except the Higgs field. Note that this characteristic is very distinct when compared with the canonical mass dimension, $\mathfrak{D} = 3/2$, of the standard model fermions, Ref. \cite{weinberg1995quantum}, such as Dirac or Majorana spinors. Self interaction is also possible for MDO field, which is expected for dark matter particles.

Some works that focused on understanding the interaction of the MDO fermion with gravity have been studied and deserve to be highlighted, such as in quantum field theory in curved spaces \cite{Pereira:2016eez} and in the covariant formulation, via gravitons interactions \cite{BuenoRogerio:2019zvz}. Now we propose to investigate the possible interaction between MDO fermions and gravity, through the canonical (Hamiltonian) formulation. Our motivation is the fact that the canonical approach was born as an attempt to build a quantum theory in which the metric functions are exempt from background perturbation, unlike the covariant perturbative formulation. 

The paper is organized as follows: in the Section \eqref{S2}, we extend the gravitational action developed by Palatini-Holst, incorporating fermionic matter with mass dimension one and spin-$1/2$, through the tetrad-ADM formalism. In Section \eqref{S3}, we investigate the classical Hamiltonian and diffeomorphism constraints that arises from this dark matter field with gravity. As a consequence, an energy density and associated directional energy flux density emerges in conjunction with the MDO-gravity coupling, which gives a term similar to the cosmological constant. Conclusions are in Section IV.

\section{Mass dimension one fermions action coupled to canonical gravity}\label{S2}

\subsection{Notation and preliminaries informations}
\indent
\indent For the purpose of studying the connection of matter to space-time through the canonical treatment of gravity, it is necessary to deal with certain preliminary notations.

\indent The Dirac gamma matrices, ${\gamma^{\mu}}$, in the Weyl basis are expressed by \cite{Boehmer:2010ma}:
\begin{equation}\label{GCMDO1}
 \gamma^{0} = \left( \begin{array}{cc}
0_{2\times2} & 1_{2\times2}\\ 
1_{2\times2} & 0_{2\times2}\\
\end{array}\right) \quad\textnormal{and}\quad \gamma^{i} = \left( \begin{array}{cc}
                                       0_{2\times2} & -\sigma^{i}\\ 
                                       \sigma^{i} & 0_{2\times2}\\
\end{array}\right),
\end{equation}
where ${\sigma^{i}}$ are the Pauli matrices:
\begin{equation}\label{GCMDO2}
 \sigma^{1} = \left( \begin{array}{cc}
0 & 1\\ 
1 & 0\\
\end{array}\right), \quad  \sigma^{2} = \left( \begin{array}{cc}
0 & -i\\ 
i & 0\\
\end{array}\right)\quad\textnormal{and}\quad  \sigma^{3} = \left( \begin{array}{cc}
1 & 0\\ 
0 & -1\\
\end{array}\right).
\end{equation}

\indent The tetrad vector fields ${e_{\mu}^{a}(x)}$ are defined by ${e_{\mu}^{a}(x)e_{\nu}^{b}(x)\eta_{ab} = g_{\mu\nu}(x)}$, being
${g_{\mu\nu}}$ the metric of spacetime and ${\eta_{ab} = (1, -1, -1, -1)}$ the metric of a local minkowskian manifold. The matrices-${\gamma}$ in the arbitrary spacetime
are given by ${\gamma^{\mu} = e^{\mu}_{a}\gamma^{a}}$ and obey Clifford's algebra:
\begin{equation}\label{GCMDO3}
 \{\gamma^{\mu},\gamma^{\nu}\} = \gamma^{\mu}\gamma^{\nu} + \gamma^{\nu}\gamma^{\mu} = 2g^{\mu\nu}
\end{equation}
and 
\begin{equation}\label{GCMDO4}
 \gamma^{5} = i\gamma^{0}\gamma^{1}\gamma^{2}\gamma^{3} = \left( \begin{array}{cc}
1_{2\times2} & 0_{2\times2} \\ 
0_{2\times2} & -1_{2\times2} \\
\end{array}\right).
\end{equation}
\indent In Quantum Field Theory (QFT) is used, in general,
the metric notation ${(+, -, -, -)}$. On the other hand, in General Relativity (GR) and Loop Quantum Gravity (LQG) the usual notation is ${(-, +, +, +)}$. To transcribe the fermionic fields in QFT to GR/LQG, we just multiply all the Dirac matrices \eqref{GCMDO1} by
${i \in Im(\mathbb{C})}$ \cite{Bojowald:2007nu}. In this paper we will make use of this last notation. Note that in fact Clifford algebra does not change,
\begin{eqnarray}\nonumber
 &&\{\gamma^{\mu},\gamma^{\nu}\}_{(QFT)} = 2g^{\mu\nu} = 2(+,-,-,-) \longmapsto \{\gamma^{\mu},\gamma^{\nu}\}_{(GR)} = \{i\gamma^{\mu},i\gamma^{\nu}\} = i^{2}\{\gamma^{\mu},\gamma^{\nu}\}_{(QFT)} = i^{2}2g^{\mu\nu}\Leftrightarrow\\
 &&\{\gamma^{\mu},\gamma^{\nu}\}_{(GR)} = -2g^{\mu\nu} = 2(-,+,+,+).\nonumber
\end{eqnarray}

\indent The set of the spinors associated with the construction of the spin ${1/2}$ MDO field is represented by eigenstates of the charge conjugation operator ${\mathcal{C}}$ with two possible eigenvalues, namely ${(\pm1)}$. Thus, ${\lambda(\boldsymbol{p},\sigma) = [\lambda^{S}_\alpha(\boldsymbol{p},\sigma), \lambda^{A}_\alpha(\boldsymbol{p},\sigma)]}$ are the eigenstates self-conjugated and anti self-conjugated associated with eigenvalues ${(+1, -1)}$, respectively, properly designated with its momentum $\boldsymbol{p}$ and spin states $\sigma$. In addition, each of these objects has a dual helicity ${\alpha = \{\pm, \mp\}}$ that emerges from its structures, as demonstrated in \cite{Ahluwalia:2004ab}, and therefore may be compactly expressed by four spinors with notation ${\lambda(\boldsymbol{p}) = [\lambda^{S}_{\{\pm, \mp\}}(\boldsymbol{p}), \lambda^{A}_{\{\pm, \mp\}}(\boldsymbol{p})]}$.

\indent In this manuscript, we will work with the representation of classical wave functions in momentum configuration space, being $\lambda(x)$ a classical field with $\lambda^{S,A}(p^{\mu})$ as its Fourier coefficients \cite{Ahluwalia:2016rwl}, namely:
\begin{equation}\label{neweq1}
    \lambda^{S}(x) = N_1\int d^{4}p\lambda^{S}(p^{\mu})e^{-ip\cdot x}\quad\textnormal{and}\quad\lambda^{A}(x) = N_2\int d^{4}p\lambda^{A}(p^{\mu})e^{+ip\cdot x},
\end{equation}
for self-conjugated $\lambda^{S}(p^{\mu})$ and anti self-conjugated $\lambda^{A}(p^{\mu})$ spinors, respectively. Here, we assume $N_1$ and $N_2$ as normalizations of wave functions that propagate in an arbitrary four-dimensional pseudo-Riemannian manifold  $\mathcal{M}$. This is a classic procedure for accommodating waves functions\footnote{For Dirac spinors, for instance, plane waves are described by $\psi(x) = u(\boldsymbol{p})e^{-ip\cdot x}$ and $\psi(x) = v(\boldsymbol{p})e^{+ip\cdot x}$, in Minkowskian spacetime, representing particles and antiparticles, respectively \cite{peskin1995quantum}.}. Thus, from now on, we will denote $\lambda(x)$ for one of the classical fields of (\ref{neweq1}).

It is important to emphasize that the decomposition used in the equation \eqref{neweq1} is not the same for the treatment of the expansion of MDO quantum fields, denoted by $\mathfrak{f}(x)$ and $\stackrel{\neg}{\mathfrak{f}}(x)$ in Ref. \cite{Ahluwalia:2016rwl}, whose expansion coefficients are the creation and annihilation operators of a one particle state, i.e., $\hat{a}^{\dagger}(\boldsymbol{p},\sigma)|0\rangle = |\boldsymbol{p},\sigma\rangle$ and $\hat{a}(\boldsymbol{p},\sigma)|0\rangle = |\boldsymbol{0},0\rangle$ $\forall~|\boldsymbol{p},\sigma\rangle \in \mathcal{H}$, where $\mathcal{H}$ is the Hilbert space. The reason for this is due to the fact that we are using the classical treatment in this work.

\indent In curved space-time, the covariant derivatives acting on the usual ${\lambda(x)}$ and dual ${\stackrel{\neg}{\lambda}(x)}$ classical MDO fields are defined by \cite{Boehmer:2010ma}: 
\begin{eqnarray}\label{GCMDO5}
 \nabla_{\mu}\lambda \equiv \partial_{\mu}\lambda - \Gamma_{\mu}\lambda\quad\textnormal{and}\quad \nabla_{\mu}\stackrel{\neg}{\lambda} \equiv \partial_{\mu}\stackrel{\neg}{\lambda} + \stackrel{\neg}{\lambda}\Gamma_{\mu},
\end{eqnarray}
and the spin connection
${\Gamma_{\mu}}$ as well as the commutation of two covariant derivatives acting on the field are expressed as (see Appendix \eqref{A}):
\begin{equation}\label{GCMDO6}
 \Gamma_{\mu} = \frac{i}{4}\omega_{\mu}^{~ab}\sigma_{ab}\quad\textnormal{and}\quad [\nabla_{\mu}, \nabla_{\nu}]\lambda = -\frac{i}{4}F_{\mu\nu}^{~~ab}\sigma_{ab}\lambda,
\end{equation}
where ${\omega_{\mu}^{~ab}\sigma_{ab} = e_{\nu}^{a}\partial_{\mu}e^{\nu b}+e_{\nu}^{a}e^{\rho b}\Gamma_{~\mu\rho}^{\nu}}$ is the connection written via tetrad fields and the  Levi-Civita connection, with spacetime (external indices ${\mu}$) and of Lorentz (internal indices ${a}$) indices.
${F_{\mu\nu}^{~~ab}}$ = ${\partial_{\mu}\omega_{\nu}^{~ab}-\partial_{\nu}\omega_{\mu}^{~ab}+\omega_{\mu}^{~ac}\omega_{\nu c}^{~~ b}-\omega_{\nu}^{~ac}\omega_{\mu c}^{~b}}$
is the ``curvature'' due to the spin connection on the spinor ${\lambda}$, with similar function to the Riemann curvature, coming from the related connection, for tensors in curved spaces, and ${\sigma_{ab} = i/2[\gamma_{a},\gamma_{b}]}$.\\

\indent Recently, Ahluwalia demonstrated that the dual spinor fields associated with the mass dimension one fermion guarantee
local symmetry of Lorentz, \cite{Ahluwalia:2016rwl}, allowing to obtain the self-conjugated and anti-self-conjugated duals as
\begin{equation}\label{GCMDO7}
 \stackrel{\neg}{\lambda}^{S}_{\alpha} = \widetilde{\lambda}^{S}_{\alpha}\mathcal{A} \quad\textnormal{and}\quad\stackrel{\neg}{\lambda}^{A}_{\alpha} = \widetilde{\lambda}^{A}_{\alpha}\mathcal{B},
\end{equation}
with ${\widetilde{\lambda}_{\alpha}(p^{\mu})\stackrel{def}{=}[\Xi(p^{\mu})\lambda_{\alpha}(p^{\mu})]^{\dagger}\gamma^{0}}$, where the ``Dirac-type'' operator is denoted as  
${\Xi(p^{\mu}) = (\mathcal{G}(p^{\mu})/m)\gamma_{\mu}p^{\mu}}$, in which the matrix ${\mathcal{G}}$ is explicitly given by
\begin{equation}\label{GCMDO8}
 {G}(p^{\mu})\stackrel{def}{=}\left( \begin{array}{cccc}
0 & 0 & 0 & -ie^{-i\varphi} \\ 
0 & 0 & ie^{i\varphi} & 0\\
0 & -ie^{-i\varphi} & 0 & 0\\
ie^{i\varphi} & 0 & 0 & 0
\end{array}\right).
\end{equation}
Assuming that such objects were constructed via Weyl's left hand spinors with the momentum in spherical coordinates:
${p^{\mu} = (E, p\sin\theta \cos\varphi, p\sin\theta \sin\varphi, p\cos\theta)}$ and ${p = |\vec{p}|}$ \cite{Ahluwalia:2004ab, Ahluwalia:2016rwl}.
The operators ${\mathcal{A}}$ and ${\mathcal{B}}$ guarantee the locality of the theory when the spin sums,
${\sum\limits_{\alpha}\lambda^{S/A}_{\alpha}\stackrel{\neg}{\lambda}^{S/A}_{\alpha}}$ \cite{Ahluwalia:2016rwl}, is obtained using the so-called $\tau-$\textit{deformation}.  

\subsection{Building the MDO fermion action in canonical gravity}
\indent 
\indent The Einstein-Hilbert action \cite{Feynman1995} with the MDO fermion action (dark matter) in curved space-time
\cite{Boehmer:2010ma, daSilva:2014kfa, Pereira:2016eez} is given by:
\begin{eqnarray}\label{GCMDO9}
 \mathcal{S} = \mathcal{S}_{EH} + \mathcal{S}_{MDO} = \frac{1}{16\pi G}\int_{\mathcal{M}}d^{4}x\sqrt{-g}R + \frac{1}{2}\int_{\mathcal{M}}d^{4}x\sqrt{-g}[g^{\mu\nu}(\nabla_{\mu}\stackrel{\neg}{\lambda}\nabla_{\nu}\lambda)-m^{2}\stackrel{\neg}{\lambda}\lambda - \xi R\stackrel{\neg}{\lambda}\lambda],
\end{eqnarray}
where ${g}$ is the determinant of metric ${g_{\mu\nu}}$, ${R}$ is the Ricci scalar curvature and ${m}$ is the mass 
associated to ${\lambda}$, characterizing the term of self-interaction ${\stackrel{\neg}{\lambda}\lambda}$. ${\xi}$ is a coupling constant between the MDO field and the gravitational field. The boundary term ${\partial \mathcal{M}}$ of the manifold ${\mathcal{M}}$ is not being considered. Explicitly, the quadratic kinetic term of MDO field is described by:
\begin{eqnarray}\label{GCMDO10}
 \nabla_{\mu}\stackrel{\neg}{\lambda}\nabla_{\nu}\lambda &=& (\partial_{\mu}\stackrel{\neg}{\lambda} + \stackrel{\neg}{\lambda}\Gamma_{\mu})(\partial_{\nu}\lambda - \Gamma_{\nu}\lambda) = \partial_{\mu}\stackrel{\neg}{\lambda}\partial_{\nu}\lambda - (\partial_{\mu}\stackrel{\neg}{\lambda})\Gamma_{\nu}\lambda + \stackrel{\neg}{\lambda}\Gamma_{\mu}(\partial_{\nu}\lambda) - \stackrel{\neg}{\lambda}\Gamma_{\mu}\Gamma_{\nu}\lambda\nonumber\\
                                                         &=& \partial_{\mu}\stackrel{\neg}{\lambda}\partial_{\nu}\lambda - (\partial_{\mu}\stackrel{\neg}{\lambda})\left[\frac{i}{4}\omega_{\nu}^{~cd}\sigma_{cd}\lambda\right] + \stackrel{\neg}{\lambda}\left[\frac{i}{4}\omega_{\mu}^{~ab}\sigma_{ab}\right](\partial_{\nu}\lambda) - \stackrel{\neg}{\lambda}\left[\frac{i}{4}\frac{i}{4}\omega_{\mu}^{~ab}\omega_{\nu}^{~cd}\sigma_{ab}\sigma_{cd}\lambda\right].
\end{eqnarray}\\
\indent Rewriting the action \eqref{GCMDO9} in terms of the tetrad fields, assuming the Einstein-Cartan formalism, considering the gravitational action
${\mathcal{S}_{EH}}$ equivalent to the action of Palatini-Holst \cite{Holst:1995pc} and using the spin connection \eqref{GCMDO5},
we have (see Appendix \eqref{B}):
\begin{eqnarray}\label{GCMDO11}
 \mathcal{S} &=& \frac{1}{16\pi G}\int_{\mathcal{M}}d^{4}x(ee^{\mu}_{I}e^{\nu}_{J}P^{IJ}_{~~~KL}F_{\mu\nu}^{~~~KL}(\omega))(1-8\pi G\xi\stackrel{\neg}{\lambda}\lambda)\nonumber\\
             &+&\frac{1}{2}\int_{\mathcal{M}}d^{4}x(e)[(e^{\mu}_{I}e^{\nu}_{J}\eta^{IJ}\partial_{\mu}\stackrel{\neg}{\lambda}\partial_{\nu}\lambda) - m^{2}\stackrel{\neg}{\lambda}\lambda]\nonumber\\
             &+&\frac{i}{8}\int_{\mathcal{M}}d^{4}x(ee^{\mu}_{I}e^{\nu}_{J}\eta^{IJ})[\stackrel{\neg}{\lambda}\omega_{\mu}^{~IJ}\sigma_{IJ}(\partial_{\nu}\lambda) - (\partial_{\mu}\stackrel{\neg}{\lambda})\omega_{\nu}^{~MN}\sigma_{MN}\lambda - i/4\stackrel{\neg}{\lambda}(\omega_{\mu}^{~IJ}\omega_{\nu}^{~MN}\sigma_{IJ}\sigma_{MN})\lambda],
\end{eqnarray}
where the curvature ${F_{\mu\nu}^{~~~KL}(\omega) = 2\partial_{[\mu}\omega_{\nu]}^{~IJ}+[\omega_{\mu},\omega_{\nu}]}$ is described in terms
of the Lorentz connection ${\omega_{\mu}^{~IJ}}$. The term ${P^{IJ}_{~~~KL}}$ of Holst action, as well as its inverse, in the compact form \cite{Bojowald:2007nu}, is written as:
\begin{eqnarray}\label{GCMDO12}
 P^{IJ}_{~~~KL} = \delta_{K}^{[I}\delta_{L}^{J]} - \frac{1}{\gamma}\frac{\epsilon^{IJ}_{~~~KL}}{2}\quad\textnormal{and}\quad P^{-1~KL}_{IJ} = \frac{\gamma^{2}}{\gamma^{2}+1}\left(\delta_{I}^{[K}\delta_{J}^{L]} + \frac{1}{\gamma}\frac{\epsilon^{~~~ KL}_{IJ}}{2}\right), 
\end{eqnarray}
where ${\gamma}$ (it can not to be confused with Dirac matrices ${\gamma^{\mu}}$) is the parameter of Barbero-Immirzi and ${\epsilon^{IJ}_{\quad KL}}$ is a tensor completely antisymmetric. It is important to clarify that the coupling term between the gravitational field and the fermion were merged in the Palatini-Holst action in order to propose ${\xi = 2G^{-1}\rho_{vac}/(R\stackrel{\neg}{\lambda}\lambda)}$ as being dependent on purely scalar factors and, therefore, invariant. Such term assumes a similar form as a time-varying cosmological model $\Lambda(t)$ \cite{Peebles:1987ek, Xu:2009mb, Alfedeel:2020lin}, for the cosmological constant ${\Lambda=8\pi\rho_{vac}}$, with ${\rho_{vac}=5,96\times10^{-27}kg/m^{3}}$ being the vacuum density \cite{Ade:2015xua}. It is also valid for scalar-tensor theories à la Brans-Dick \cite{brans1961mach, Molinari:2022rff, M:2022xcp}. This proposal is constructed for reference frames in which the Holst term recovers the exact scalar Ricci curvature, $R$, originally present in the Einstein-Hilbert action. Furthermore, it is noted that the matter part, in our case, is written as a sum of two parts in the action of the MDO field: the first part resembles the dynamics of a massive scalar field and the second part carries the quadratic information of the pure fermionic portion of the MDO fermion with spin connection (the last term integrated over the ${\mathcal{M}}$ manifold of action \eqref{GCMDO11}). This is due to the Klein-Gordon dynamic nature of energy conservation and the structure of this fermion, respectively.\\

\indent A canonical (Hamiltonian) formalism is built from a Legendre transformation in action. For this, it will be necessary to transcribe
the action \eqref{GCMDO11} from manifold ${\mathcal{M}}$, provided with a metric, to a family of constant foliations ${\Sigma_{t}}$, and determined by the function of time ${t}$, which are embedded in the manifold ${(\mathcal{M}, g_{ab})\longmapsto (\Sigma_{t}, t)\equiv\Sigma_{t}\times\mathbb{R}\subseteq\mathcal{M}}$,
as originally prescribed by \cite{Arnowitt:1959ah}, known as ADM (Arnowitt-Deser-Misner) formalism. Instead of working with spacetime tensors, we will use the description of how
a vector field ${t^{\mu}}$ evolves from a hypersurface ${\Sigma_{t}}$ to another, ${\Sigma_{t+\delta t}}$, in direction to the future, since it
is using a Lorentzian metric signature. Following the same approach used in \cite{Bojowald:2007nu}, the four-vector, in spacetime, ${t^{\mu}}$ is
\begin{equation}\label{GCMDO13}
 t^{\mu} = Nn^{\mu} + N^{\mu},
\end{equation}
where ${n^{\mu}}$ is the vector normal to the hypersurface ${\Sigma_{t}}$, ${N^{\mu}}$ is the \textit{shift} vector tangent to the surface ${\Sigma_{t}}$ and
${N}$ is the \textit{lapse} function which dictates the temporal transition from one hypersurface to another, therefore, ${N^{\mu}n_{\mu} = 0}$. 
The spatial evolution of ${t^{a}}$ must obey ${t^{a}\nabla_{a}t = 1}$, being ${a}$, ${b}$, ${c}$,... spatial tensor indices. The metric ${g_{\mu\nu}}$ is described, therefore, as 
\begin{equation}\label{GCMDO14}
 g_{\mu\nu} = q_{\mu\nu} - n_{\mu}n_{\nu}.
\end{equation}
Since we are using the tetrad formalism in addition to the ADM formalism, it is necessary to perform a gauge fixation in internal vector fields of the tetrad, thus it can be decomposed into internal time units, as a vector and a \textit{triad} (spatial part of the tetrad). Setting the internal vector field
as a constant ${n_{I} = -\delta_{I}^{0}}$, where ${n^{I}n_{I} = 1}$, we get:
\begin{eqnarray}\label{GCMDO15}
 &&n^{a} = n^{I}e_{I}^{a}\quad\textnormal{and}\quad e_{I}^{a} = \varepsilon_{I}^{a} - n^{a}n_{I},
\end{eqnarray}
where ${n^{a}}$ is the unit normal to foliation, ${\varepsilon_{I}^{a}n_{a} = \varepsilon_{I}^{a}n^{I} = 0}$, with 
${\varepsilon_{I}^{a}:=}$ \textit{triad}. From equation \eqref{GCMDO13}, one has
\begin{equation}\label{GCMDO16}
 n^{a} = N^{-1}(t^{a} - N^{a}),
\end{equation}
for the normal and tangential projection of \eqref{GCMDO14} in ${\Sigma_{t}}$.\\
\indent Thus, using ADM  formalism decomposition and tetrad fields, the action of MDO matter with the Palatini-Holst gravitation is described as (see Appendix \eqref{C}):
\begin{eqnarray}\label{GCMDO17}
 \mathcal{S}[\varepsilon, \omega, \lambda] &=& \mathcal{S}_{PH,\xi} + \mathcal{S}_{\lambda, \stackrel{\neg}{\lambda}_{S}} + \mathcal{S}_{\lambda, \stackrel{\neg}{\lambda}_{F}}\nonumber\\
                                 &=&\frac{1}{16\pi G}\int_{\mathbb{R}}dt\int_{\Sigma}d^{3}xN\sqrt{q}\Omega_{IJ}^{ab}P^{IJ}_{~~~KL}F_{ab}^{~~~KL}(\omega)(1-8\pi G\xi\stackrel{\neg}{\lambda}\lambda)\nonumber\\
                                 &&+\frac{1}{2}\int_{\mathbb{R}}dt\int_{\Sigma}d^{3}xN\sqrt{q}[\Omega_{IJ}^{ab}\eta^{IJ}(\partial_{a}\stackrel{\neg}{\lambda}\partial_{b}\lambda)-m^{2}\stackrel{\neg}{\lambda}\lambda]\nonumber\\
                                 &&-\frac{1}{8}\int_{\mathbb{R}}dt\int_{\Sigma}d^{3}xN\sqrt{q}\Omega_{IJ}^{ab}\eta^{IJ}\{i/2(\stackrel{\neg}{\lambda}\omega_{a}^{IJ}[\gamma_{I},\gamma_{J}](\partial_{b}\lambda)\nonumber\\
                                 &&- (\partial_{a}\stackrel{\neg}{\lambda})\omega_{b}^{MN}[\gamma_{M},\gamma_{N}]\lambda)+i/16\stackrel{\neg}{\lambda}(\omega_{a}^{IJ}\omega_{b}^{MN}[\gamma_{I},\gamma_{J}][\gamma_{M},\gamma_{N}])\lambda\},
\end{eqnarray}
where $\Omega_{IJ}^{ab} = (\varepsilon_{I}^{a}\varepsilon_{J}^{b}-2N^{-1}n_{I}t^{a}\varepsilon_{J}^{b} + 2N^{-1}N^{a}n_{I}\varepsilon_{J}^{b})$. The subscripts in the action of matter, $S_{\lambda, \stackrel{\neg}{\lambda}_{S}}$ and $S_{\lambda, \stackrel{\neg}{\lambda}_{F}}$, are to emphasize the difference between two terms: a part with characteristic similar to the Klein-Gordon dynamics (scalar - S)  and other with pure fermionic quadratic dynamics (fermionic - F), respectively, due to the aforementioned nature of this fermion.

The equation \eqref{GCMDO17}, for now, presents the Lagrangian density of a dark fermionic matter coupled to gravity with the contribution of dark energy for models $\Lambda(t)$ or Brans-Dick theory, admitting ${\xi = 2G^{-1}\rho_{vac}/(R\stackrel{\neg}{\lambda}\lambda)}$. 

\section{Hamiltonian constraint between MDO and gravity}\label{S3}

The action \eqref{GCMDO17}, and its Lagrangian density, is the most complete way of treating the Hamiltonian formalism in terms of the classical matter field of interest. In the scope of this work, the scenario without torsion and with the Palatini-Holst dual term  recovers the usual Ricci scalar structure. Under a family of foliations $\Sigma$ in the manifold and with the aid of the expression \eqref{GCMDO9}, one has
\begin{eqnarray}\label{GCMDO18}
    \mathcal{S} &=& \frac{1}{16\pi G}\int_{\mathbb{R}}dt\int_{\Sigma}d^{3}x N(G^{abcd}K_{ab}K_{cd} + \sqrt{q}^{(3)}R)[1-8\pi G\xi\stackrel{\neg}{\lambda}\lambda]\nonumber\\
    &+&\frac{1}{2}\int_{\mathbb{R}}dt\int_{\Sigma}d^{3}xN\sqrt{q}[(q^{\mu\nu}+n^{\mu}n^{\nu})(q_{\mu}^{\alpha}D_{\alpha}\stackrel{\neg}{\lambda}q_{\nu}^{\beta}D_{\beta}\lambda)-m^{2}\stackrel{\neg}{\lambda}\lambda],
\end{eqnarray}
where $D_{\alpha} = q^{\mu}_{\alpha}\nabla_{\mu}$ is the projection of the covariant derivative operator on $\mathcal{M}$ onto the hypersurface $\Sigma_t$ and the pair $(G^{abcd},~G_{abcd})$ is the DeWitt metric, $K_{ab} = \Gamma_{ab}^{\mu}n_{\mu} = -\Gamma^{0}_{ab}/\sqrt{-g^{00}}$ is a symmetrical “velocity” associated with the three-metric $q_{ab}$ \cite{kiefer2007quantum}, expressed in terms of the spatial components of the equation \eqref{GCMDO14}. Thus, both are written respectively as:
\begin{equation}\label{GCMDO19}
    G^{abcd} = \frac{\sqrt{q}}{2}(q^{ac}q^{bd}+q^{ad}q^{bc}-2q^{ab}q^{cd})\quad\textnormal{and}\quad G_{abcd} = \frac{1}{2\sqrt{q}}(q_{ac}q_{bd}+q_{ad}q_{bc}-2q_{ab}q_{cd}),
\end{equation}
and 
\begin{equation}\label{GCMD20}
    K_{\mu\nu} \equiv q_{\mu}^{~\rho}q_{\nu}^{~\sigma}\nabla_{\rho}n_{\sigma}= q_{\mu}^{~\rho}\nabla_{\rho}n_{\nu} = K_{\nu\mu},
\end{equation}
which characterizes a purely special form, as well as $q_{\mu\nu}$, furthermore $K_{\mu\nu}$ is orthogonal to $n^{\mu}$ $(K_{\mu\nu}n^{\mu} = K_{ \mu\nu}n^{\nu} = 0)$.

In order to seek a canonical approach that couples this dark matter candidate with gravity, we will start by choosing a configuration variable and defining the momentum for the MDO $('p=\partial L/\partial\dot{q}')$, in order to ensure the Legendre transformation:
\begin{equation}\label{GCMD21}
     H_{\lambda, \stackrel{\neg}{\lambda}}=p_{\lambda}\dot{\lambda}+p_{\stackrel{\neg}{\lambda}}\dot{\stackrel{\neg}{\lambda}}-\mathcal{L}_{\lambda, \stackrel{\neg}{\lambda}}.
 \end{equation}
Starting from the explicit Lagrangian density of the MDO fermion,
\begin{eqnarray}\label{GCMD22}
\mathcal{L}_{\lambda, \stackrel{\neg}{\lambda}} = \frac{1}{2}\sqrt{-g}[\partial^{\nu}\stackrel{\neg}{\lambda}\partial_{\nu}\lambda - \partial^{\nu}\stackrel{\neg}{\lambda}\Gamma_{\nu}\lambda+\stackrel{\neg}{\lambda}\Gamma^{\nu}\partial_{\nu}\lambda - \stackrel{\neg}{\lambda}\Gamma^{\nu}\Gamma_{\nu}\lambda - m^{2}\stackrel{\neg}{\lambda}\lambda],
\end{eqnarray}
we can get the following operators
\begin{eqnarray}\label{GCMD23}
\frac{\partial\mathcal{L}_{\lambda, \stackrel{\neg}{\lambda}}}{\partial(\partial_{\alpha}\lambda)} = \frac{\sqrt{-g}}{2}(\partial^{\nu}\stackrel{\neg}{\lambda}\delta_{\nu}^{\alpha}+\stackrel{\neg}{\lambda}\Gamma^{\nu}\delta_{\nu}^{\alpha})\quad\textnormal{and}\quad\frac{\partial\mathcal{L}_{\lambda, \stackrel{\neg}{\lambda}}}{\partial(\partial_{\alpha}\stackrel{\neg}{\lambda})} = \frac{\sqrt{-g}}{2}(\delta_{\beta}^{\alpha}\partial^{\beta}\lambda - \delta_{\beta}^{\alpha}\Gamma^{\beta}\lambda)
\end{eqnarray}
allowing us to construct their conjugate momentum as
\begin{eqnarray}
    p_{\lambda} = \frac{\partial\mathcal{L}_{\lambda, \stackrel{\neg}{\lambda}}}{\partial(\partial_{0}\lambda)} = \frac{\sqrt{-g}}{2}(\partial^{0}\stackrel{\neg}{\lambda} + \stackrel{\neg}{\lambda}\Gamma^{0})\quad\textnormal{and}\quad
    p_{\stackrel{\neg}{\lambda}} = \frac{\partial\mathcal{L}_{\lambda, \stackrel{\neg}{\lambda}}}{\partial(\partial_{0}\stackrel{\neg}{\lambda})} = \frac{\sqrt{-g}}{2}(\partial^{0}\lambda - \Gamma^{0}\lambda)\label{GCMD25}.
\end{eqnarray}
Then, the respective “velocity” of the field in terms of momentum is determined, with the help of \eqref{GCMD25}, as
\begin{eqnarray}
    \dot{\lambda}= \nabla_{0}\lambda = (\partial_{0}\lambda-\Gamma_{0}\lambda) = \frac{2}{\sqrt{-g}}g_{00}p_{\stackrel{\neg}{\lambda}}\quad\textnormal{and}\quad\dot{\stackrel{\neg}{\lambda}} = \nabla_{0}\stackrel{\neg}{\lambda} = (\partial_{0}\stackrel{\neg}{\lambda}+\stackrel{\neg}{\lambda}\Gamma_{0}) = \frac{2}{\sqrt{-g}}g_{00}p_{\lambda}.\label{GCMD26}
\end{eqnarray}
We also note that the Lagrangian density \eqref{GCMD22} can be written more explicitly as

\begin{eqnarray}\label{GCMD27}
\mathcal{L}_{\lambda,\stackrel{\neg}{\lambda}} &=& \frac{1}{2}\sqrt{-g}\underbrace{[\partial^{0}\stackrel{\neg}{\lambda}\partial_{0}\lambda - \partial^{0}\stackrel{\neg}{\lambda}\Gamma_{0}\lambda+\stackrel{\neg}{\lambda}\Gamma^{0}\partial_{0}\lambda-\stackrel{\neg}{\lambda}\Gamma^{0}\Gamma_{0}\lambda]}_{\mathbb{A}}-\frac{1}{2}\sqrt{-g}[\partial^{i}\stackrel{\neg}{\lambda}\partial_{i}\lambda + m^{2}\stackrel{\neg}{\lambda}\lambda]\nonumber\\
&-&\frac{1}{2}\sqrt{-g}[\stackrel{\neg}{\lambda}\Gamma^{i}\partial_{i}\lambda-\partial^{i}\stackrel{\neg}{\lambda}\Gamma_{i}\lambda-\stackrel{\neg}{\lambda}\Gamma^{i}\Gamma_{i}\lambda],
\end{eqnarray}
where
\begin{equation}\label{GCMD28}
    \mathbb{A} = (\partial^{0}\stackrel{\neg}{\lambda}+\stackrel{\neg}{\lambda}\Gamma^{0})(\partial_{0}\lambda-\Gamma_{0}\lambda) = \frac{2}{\sqrt{-g}}\frac{2g_{00}}{\sqrt{-g}}(p_{\lambda}p_{\stackrel{\neg}{\lambda}}).
\end{equation}
Thus, one has
\begin{eqnarray}\label{GCMD29}
\mathcal{L}_{\lambda,\stackrel{\neg}{\lambda}} = \frac{2}{\sqrt{-g}}g_{00}(p_{\lambda}p_{\stackrel{\neg}{\lambda}})-\frac{1}{2}\sqrt{-g}[\partial^{i}\stackrel{\neg}{\lambda}\partial_{i}\lambda + m^{2}\stackrel{\neg}{\lambda}\lambda]-\frac{1}{2}\sqrt{-g}[\stackrel{\neg}{\lambda}\Gamma^{i}\partial_{i}\lambda-\partial^{i}\stackrel{\neg}{\lambda}\Gamma_{i}\lambda-\stackrel{\neg}{\lambda}\Gamma^{i}\Gamma_{i}\lambda].
\end{eqnarray}
With the Legendre transformation performed on $\mathcal{L}_{\lambda,\stackrel{\neg}{\lambda}}$, in the expression \eqref{GCMD29}, we develop the Hamiltonian density of the dark fermion in curved space,
\begin{eqnarray}\label{GCMD30}
\mathcal{H}_{\lambda,\stackrel{\neg}{\lambda}} = p_{\lambda}\dot{\lambda} + p_{\stackrel{\neg}{\lambda}}\dot{\stackrel{\neg}{\lambda}} - \mathcal{L}_{\lambda,\stackrel{\neg}{\lambda}} =
\frac{2}{\sqrt{-g}}g_{00}(p_{\stackrel{\neg}{\lambda}}p_{\lambda})+\frac{1}{2}\sqrt{-g}(\partial^{i}\stackrel{\neg}{\lambda}\partial_{i}\lambda + m^{2}\stackrel{\neg}{\lambda}\lambda)+\frac{1}{2}\sqrt{-g}[\stackrel{\neg}{\lambda}\Gamma^{i}\partial_{i}\lambda-\partial^{i}\stackrel{\neg}{\lambda}\Gamma_{i}\lambda-\stackrel{\neg}{\lambda}\Gamma^{i}\Gamma_{i}\lambda),
\end{eqnarray}
which in turn allows us to finally describe the Hamiltonian of the MDO fermion in a foliation scenario, $\Sigma_{t}$, à lá \textit{ADM}, in the form:
\begin{eqnarray}\label{GCMD31}
H_{\lambda,\stackrel{\neg}{\lambda}} &=& \int_{\Sigma} d^{3}x\mathcal{H}_{\lambda,\stackrel{\neg}{\lambda}}\nonumber\\
&=&\int_{\Sigma}d^{3}xN\left(\frac{2p_{\stackrel{\neg}{\lambda}}p_{\lambda}}{\sqrt{q}}+\frac{\sqrt{q}}{2}(q^{ab}\partial_{a}\stackrel{\neg}{\lambda}\partial_{b}\lambda+m^{2}\stackrel{\neg}{\lambda}\lambda)\right)\nonumber\\
&+&\int_{\Sigma}d^{3}xN^{a}\left(\frac{1}{2}q_{a}^{b}[\stackrel{\neg}{\lambda}\Gamma_{b}\partial_{c}\lambda-\partial_{b}\stackrel{\neg}{\lambda}\Gamma_{c}\lambda-\stackrel{\neg}{\lambda}\Gamma_{b}\Gamma_{c}\lambda]n^{c}\right)\nonumber\\
&\equiv&\int_{\Sigma}d^{3}x(N\mathcal{H}_{\perp,\lambda,\stackrel{\neg}{\lambda}}+N^{a}\mathcal{H}_{a,\lambda,\stackrel{\neg}{\lambda}}),
\end{eqnarray}
noting that the transformations $\sqrt{-g}\mapsto N\sqrt{q}$ and $\sqrt{-g}g^{ij}A_{j}B_{i}\mapsto N^{a}q_{ a}^{b}(A_{b}B_{c})n^{c}$ were performed to map $\mathcal{M}\longrightarrow\mathbb{R}\times\Sigma$.

The equation \eqref{GCMD31} allows us to visualize $\mathcal{H}_{\perp,\lambda,\stackrel{\neg}{\lambda}}$ as the energy density of the MDO field $(T_{00} = \rho_{\lambda,\stackrel{\neg}{\lambda}})$, which “flows” through the \textit{lapse} function, and consequently orthogonal to $ \Sigma$, at a given fixed instant of time $t$. The term $\mathcal{H}_{a,\lambda,\stackrel{\neg}{\lambda}}$ can be interpreted as a "flux" $(J_{a, \lambda,\stackrel{\neg }{\lambda}} \equiv q_{a}^{~b}T_{bc}n^{c})$ associated with the MDO fermions, which moves through the shift vector  $N^{a}$ tangential to the hypersurface $\Sigma_{t}$. It is notable that this directional energy density flux depends exclusively on the spatial derivative and the spin connection of the mass-dimension-one field.

We can clarify the physical meaning of the Hamiltonian and diffeomorphism constraints for gravity in this scenery. It is important to note that in the absence of matter, the structures of gravitational constraints, denoted in the expression below by $\mathcal{H}_{\perp,~g}$ (Hamiltonian) and $\mathcal{H}_{a,~g}$ (diffeomorphism) respectively, are well established in the literature, according to \cite{Arnowitt:1959ah, kiefer2007quantum}. Now, we obtained a result with geometry of space-time and a peculiar material content, i.e., gravity and the MDO fermion, as developed from the action \eqref{GCMDO18}. Therefore, one has:
\begin{eqnarray}
\mathcal{H}_{\perp} &=& \mathcal{H}_{\perp,~g} + \mathcal{H}_{\perp,\lambda,\stackrel{\neg}{\lambda}} = \left(16\pi G G_{abcd}p^{ab}p^{cd} - \frac{\sqrt{q}}{16\pi G}^{(3)}R\right)[1-8\pi G\xi\stackrel{\neg}{\lambda}\lambda] + \rho_{\lambda,\stackrel{\neg}{\lambda}} \approx0\quad\textnormal{and}\quad\label{GCMD32}\\
\mathcal{H}_{a} &=& \mathcal{H}_{a,~g} + \mathcal{H}_{a,\lambda,\stackrel{\neg}{\lambda}} = -2D_{b}p_{a}^{~b}[1-8\pi G\xi\stackrel{\neg}{\lambda}\lambda] + J_{a, \lambda,\stackrel{\neg}{\lambda}}\approx 0\label{GCMD33},
\end{eqnarray}
or even more explicitly, using the equation \eqref{GCMD30},
\begin{eqnarray}
\mathcal{H}_{\perp} &=& \left(16\pi G G_{abcd}p^{ab}p^{cd} - \frac{\sqrt{q}}{16\pi G}^{(3)}R\right)[1-8\pi G\xi\stackrel{\neg}{\lambda}\lambda]+ \left(\frac{2p_{\stackrel{\neg}{\lambda}}p_{\lambda}}{\sqrt{q}}+\frac{\sqrt{q}}{2}(q^{ab}\partial_{a}\stackrel{\neg}{\lambda}\partial_{b}\lambda+m^{2}\stackrel{\neg}{\lambda}\lambda)\right) \approx0\quad\textnormal{and}\quad\label{GCMD34}\\
\mathcal{H}_{a} &=& -2D_{b}p_{a}^{~b}[1-8\pi G\xi\stackrel{\neg}{\lambda}\lambda]+ \left(\frac{1}{2}q_{a}^{b}[\stackrel{\neg}{\lambda}\Gamma_{b}\partial_{c}\lambda-\partial_{b}\stackrel{\neg}{\lambda}\Gamma_{c}\lambda-\stackrel{\neg}{\lambda}\Gamma_{b}\Gamma_{c}\lambda]n^{c}\right)\approx 0\label{GCMD35},
\end{eqnarray}
remembering that the term $-8\pi G\xi\stackrel{\neg}{\lambda}\lambda$ was merged to the Ricci scalar, in the geometric portion of the action, in order to guarantee a similar behavior to the cosmological constant in $\Lambda(t)$ models. This new term comes from the coupling constant $\xi$ between MDO and gravity, and from the orthonormal invariance $\stackrel{\neg}{\lambda}(x)\lambda(x') = \pm 2m\delta(x-x')$ for the same helicities of the fermion, in its self-conjugated $(S)$ and anti-self-conjugated $(A)$ forms, respectively.

Just as a comparative analysis, when incorporating a scalar Lagrangian density field $\mathcal{L}_{\phi} = \sqrt{-g}(-1/2g^{\mu\nu}\nabla_{\mu} \phi\nabla_{\nu}\phi - 1/2m^{2}\phi^{2})$ to same gravitational scenario \cite{kiefer2007quantum}, the constraints of scalar field $\phi$, $\mathcal{ H}_{\perp,\phi}$ and $\mathcal{H}_{a,\phi}$, in curved space-time are given by
\begin{eqnarray}\label{GCMD36}
\mathcal{H}_{\perp,\phi} = \rho_{\phi} = \left(\frac{p_{\phi}^{2}}{2\sqrt{q}}+\frac{\sqrt{q}}{2}(q^{ab}\partial_{a}\phi\partial_{b}\phi+m^{2}\phi^{2})\right)\quad\textnormal{and}\quad \mathcal{H}_{a,\phi} = J_{a,\phi} = p_{\phi}\partial_{a}\phi.
\end{eqnarray}

It is possible to notice, in \eqref{GCMD34}-\eqref{GCMD35} and \eqref{GCMD36}, that the energy density of the MDO field, ($\rho_{ \lambda,\stackrel{\neg}{\lambda}}$), closely resembles the one for scalar field ($\rho_{\phi}$), since the dynamic equation of both fields follows the  Klein Gordon equation in the same way. However, while the “current” - directional density of energy flux - in the hypersurface $\Sigma_{t}$ of the MDO field ($J_{a,\stackrel{\neg}{\lambda},\lambda }$) is a function of the spatial derivatives and spin connections, the scalar field current is also a function of the spatial derivative, but it has no spin connection term, as expected. This difference is clarified due to the spinor structure of the fermion in curved space.


\section{Conclusion}\label{S4}
\indent
\indent In this work we present a proposal for the incorporation of the mass-dimension-one fermion with gravity, under a canonical approach. We start the first section with the Hamiltonian formulation of General Relativity using the ADM formalism \cite{Arnowitt:1959ah}, which foliate a pseudo-Riemannian manifold, equipped with a physical metric $(\mathcal{M},g_{\mu \nu})$, for a globally hyperbolic spacetime, on Cauchy hypersurfaces $\Sigma_{t}$ for each fixed $t\in\mathbb{R}$, being $\mathcal{M}\simeq\mathbb{R}\times\Sigma$. In this formulation, the existence of two fundamental forms, represented by the three-metric decomposition and the extrinsic curvature $(q_{\mu\nu}, K_{\mu\nu})$, in conjunction with the projection of the covariant derivative $D_{\mu }$ in the hypersurface, are responsible for characterizing the so-called ADM action, whose Legendre transformation on its Lagrangian density results in a geometric Hamiltonian and diffeomorphism constraints associated with gravity and which establish the so-called "dynamics of the theory in the absence of matter". In the subsequent section, we effectively incorporate the MDO fermion into the Einstein-Hilbert action. Aiming on a greater completeness, via fermionic and gravitational interaction, the geometric part of the Einstein-Hilbert action was rewritten in terms of tetrads into an extension of Palatini \cite{Holst:1995pc}, which holds a dual field for the original gravitational theory. The portion associated with the coupling constant between fermion and gravity was rewritten in order to reproduce the behavior of a cosmological constant term. Performing the same foliation process, in the light of the ADM formalism, we build an action that combines dark matter MDO and gravity via Palatini-Holst in ADM-tetrad \eqref{GCMDO17}, comprising the most complete form for a Hamiltonian analysis in terms of the field of interest.

Noticing that the dual term of Palatini-Holst, when it is null in eq \eqref{GCMDO17}, recovers the original form for the Ricci scalar, we can investigate the Hamiltonian formulation developed according to the ADM formalism. For our context, we will obtain the constraints between the gravitational Hamiltonian density and the MDO fermion. After the proper Legendre transformation for the fermion  Lagrangian density \eqref{GCMD22}, in curved space, we were able to write the Hamiltonian in a foliation scenario $\Sigma_{t}$ by means of ADM formalism, which explained the Hamiltonian component $\mathcal{H }_{\perp,\lambda,\stackrel{\neg}{\lambda}}$ as the energy density of the MDO fermion $(T_{00} = \rho_{\lambda,\stackrel{ \neg}{\lambda}})$, which “flows” through the lapse function, orthogonal to $\Sigma$ at a given instant of fixed time $t$. Moreover, $\mathcal{H}_{a,\lambda,\stackrel{\neg}{\lambda}}$ can be decomposed as a directional energy flux density, which is a function of spatial derivatives and spin connections of the MDO field under the representation $(J_{a, \lambda,\stackrel{\neg}{\lambda}} \equiv q_{a}^{~b}T_{bc}n^{ c})$, which moves through the shift vector $N^{a}$ tangential to the hypersurface $\Sigma_{t}$. 

In this way, we obtained the Hamiltonian and diffeomorphic constraints due to the incorporation of the MDO matter to the action, via ADM formalism, expressed by the equations \eqref{GCMD35}-\eqref{GCMD36}, which elucidate three ingredients for canonical dynamics, namely: gravity, dark matter and dark energy. The latter one can be interpreted through the varying cosmological constant like term in $\Lambda(t)$ models, now encoded in the term $-8\pi G \xi\stackrel{\neg}{\lambda}\lambda$, where $\stackrel{\neg}{\lambda}\lambda$ is an orthonormal invariant for each and every observable, $\stackrel{\neg}{\lambda }(x)\lambda(x') = \pm 2m\delta(x-x')$, dictated for the same helicities of the fermion in its self-conjugated $(S)$ and anti-self-conjugated $(A)$ form, respectively. 

It is important to call attention for the fact that such contribution is absent in Dirac fermion sector. Looking for the action \eqref{GCMDO9} we see that $\xi$ is a dimensionless coupling, what forbids the coupling of a mass dimension 3/2 fermion to it. A Dirac fermion $\psi$ could couple to another dimensionfull constant $\zeta$ through $\zeta R \bar{\psi}\psi$, where  $\zeta$ has dimension $[Mass]^{-1}$. Thus, such effect of non-minimal coupling to gravity through a dimensionless constant is restrict to scalar fields or mass dimension one fermions.\footnote{Refs. \cite{Bojowald:2007nu,thiemann2008modern} uses $\zeta = 0$, so that such contribution for DE is absent.}

The role of this coupling due to $\xi$ can also be interpreted as a contributing part in Brans-Dicke \cite{GabrieleGionti:2020drq} scalar-tensor theories. We also notice that the energy density $\rho_{\lambda,\stackrel{\neg}{\lambda}}$ coming from $\mathcal{H}_{\perp,\lambda,\stackrel{\neg} {\lambda}}$ has a similar form to that expected for a scalar field $(\rho_{\phi})$ in the same scenario, unlike the structure observed for directional flow of energy between both matter fields, $J_{a} $. It should be noted that this construction related to the constraints was carried out on a general metric $g_{\mu\nu} = q_{\mu\nu} - n_{\mu}n_{\nu}$, thus being able to be applied to different scenarios of cosmological interests such as FLRW, black holes, wormholes among others. 


Furthermore, a possible hypothesis raised for future investigations would be to resort to Ashtekar's new canonical variables $(E_{i}^{a}(x), A_{b}^{j}(x'))$ \cite{Ashtekar:1987gu}, which presently support the bases of the modern canonical quantization theory of gravity, Loop Quantum Gravity (LQG). LQG is a quantum theory of gravitation based on a geometric formulation, whose intention is to unify Quantum Mechanics and General Relativity, incorporating the Standard Model matter to the established framework for the case of pure quantum gravity. We intend to use this proposal of gravitation directly in the action \eqref{GCMDO17} in search of new constraints between gravity-MDO, which can lead to quantization, in the future.

\section{Acknowledgement}
RdCL thank to Coordenação de Aperfeiçoamento de Pessoal de Nível Superior - Brasil (CAPES) - Finance Code 001. SHP acknowledges financial support from  {Conselho Nacional de Desenvolvimento Cient\'ifico e Tecnol\'ogico} (CNPq)  (No. 303583/2018-5 and 308469/2021-6).

\appendix

\section{Covariant derivatives commutation of MDO field}\label{A}
\indent
\indent Using the definition of the spin connection \eqref{GCMDO6} in the covariant derivative in equation \eqref{GCMDO5}, and performing the commutation of two covariant derivatives over  ${\lambda}$, has been:
\begin{eqnarray}\label{GCMDOA1}
 [\nabla_{\mu}, \nabla_{\nu}]\lambda &=& \nabla_{\mu}\nabla_{\nu}\lambda - \nabla_{\nu}\nabla_{\mu}\lambda = \nabla_{\mu}\left(\partial_{\nu}-\frac{i}{4}\omega_{\nu}^{ab}\sigma_{ab}\right)\lambda - \nabla_{\nu}\left(\partial_{\mu}-\frac{i}{4}\omega_{\mu}^{ab}\sigma_{ab}\right)\lambda\nonumber\\
                                     &=& \Big[\partial_{\mu}\left(\partial_{\nu}-\frac{i}{4}\omega_{\nu}^{ab}\sigma_{ab}\right)\lambda - \Gamma_{\mu\nu}^{\rho}\left(\partial_{\rho}-\frac{i}{4}\omega_{\rho}^{ab}\sigma_{ab}\right)\lambda - \frac{i}{4}\omega_{\mu}^{ab}\sigma_{ab}(\partial_{\nu}\lambda) + \frac{i}{4}\omega_{\mu}^{ab}\sigma_{ab}\frac{i}{4}\omega_{\nu}^{cd}\sigma_{cd}\lambda\Big]\nonumber\\
                                     &-& \Big[\partial_{\nu}\left(\partial_{\mu}-\frac{i}{4}\omega_{\mu}^{ab}\sigma_{ab}\right)\lambda - \Gamma_{\nu\mu}^{\rho}\left(\partial_{\rho}-\frac{i}{4}\omega_{\rho}^{ab}\sigma_{ab}\right)\lambda - \frac{i}{4}\omega_{\nu}^{ab}\sigma_{ab}(\partial_{\mu}\lambda) + \frac{i}{4}\omega_{\nu}^{cd}\sigma_{cd}\frac{i}{4}\omega_{\mu}^{ab}\sigma_{ab}\lambda\Big],                                     
\end{eqnarray}
being ${(\partial_{\mu}\partial_{\nu} - \partial_{\nu}\partial_{\mu})\lambda = 0}$ and assuming a torsion free scenario
${(\Gamma_{\mu\nu}^{\rho}-\Gamma_{\nu\mu}^{\rho})=0}$, we have
\begin{eqnarray}\label{GCMDOA2}
 [\nabla_{\mu}, \nabla_{\nu}]\lambda &=& -\frac{i}{4}\partial_{\mu}(\omega_{\nu}^{ab}\sigma_{ab}\lambda) - \frac{i}{4}\omega_{\mu}^{ab}\sigma_{ab}(\partial_{\nu}\lambda) +\frac{i}{4}\partial_{\nu}(\omega_{\mu}^{ab}\sigma_{ab}\lambda) + \frac{i}{4}\omega_{\nu}^{ab}\sigma_{ab}(\partial_{\mu}\lambda)\nonumber\\
                                     &&+\frac{i}{4}\frac{i}{4}(\omega_{\mu}^{ab}\sigma_{ab}\omega_{\nu}^{cd}\sigma_{cd} - \omega_{\nu}^{cd}\sigma_{cd}\omega_{\mu}^{ab}\sigma_{ab})\lambda\nonumber\\
                                     &=&-\frac{i}{4}(\partial_{\mu}\omega_{\nu}^{ab})\sigma_{ab}\lambda - \frac{i}{4}\omega_{\nu}^{ab}\sigma_{ab}(\partial_{\mu}\lambda)-\frac{i}{4}\omega_{\mu}^{ab}\sigma_{ab}(\partial_{\nu}\lambda) + \frac{i}{4}(\partial_{\nu}\omega_{\mu}^{ab})\sigma_{ab}\lambda\nonumber\\
                                     &&+\frac{i}{4}\omega_{\mu}^{ab}\sigma_{ab}(\partial_{\nu}\lambda) + \frac{i}{4}\omega_{\nu}^{ab}\sigma_{ab}(\partial_{\mu}\lambda)+\frac{i}{4}\frac{i}{4}(\omega_{\mu}^{ab}\sigma_{ab}\omega_{\nu}^{cd}\sigma_{cd} - \omega_{\nu}^{cd}\sigma_{cd}\omega_{\mu}^{ab}\sigma_{ab})\lambda\nonumber\\
                                     &=&-\frac{i}{4}(\partial_{\mu}\omega_{\nu}^{ab}-\partial_{\nu}\omega_{\mu}^{ab})\sigma_{ab}\lambda + \frac{i}{4}\frac{i}{4}\omega_{\mu}^{ab}\omega_{\nu}^{cd}(\sigma_{ab}\sigma_{cd}-\sigma_{cd}\sigma_{ab})\lambda.
\end{eqnarray}
Once ${\sigma^{ab}=i/2[\gamma^{a},\gamma^{b}]}$ and ${\{\gamma^{a},\gamma^{b}\}=2\eta^{ab}}$, immediately we have ${\gamma^{b}\gamma^{a}=2\eta^{ab}-\gamma^{a}\gamma^{b}}$ and ${\sigma^{ab} = i/2[\gamma^{a},\gamma^{b}]}$ ${=i/2(\gamma^{a}\gamma^{b}-2\eta^{ab}+\gamma^{a}\gamma^{b}) = i(\gamma^{a}\gamma^{b}-\eta^{ab})}$. Using ${\sigma^{ab}}$ in the last term of expression \eqref{GCMDOA2}; 
\begin{eqnarray}\label{GCMDOA3}
 [\nabla_{\mu}, \nabla_{\nu}]\lambda &=& -\frac{i}{4}(\partial_{\mu}\omega_{\nu}^{ab}-\partial_{\nu}\omega_{\mu}^{ab})\sigma_{ab}\lambda - \frac{i}{4}(\omega_{\mu}^{ac}\omega_{\nu c}^{\quad b}-\omega_{\nu}^{ac}\omega_{\mu c}^{\quad b})\sigma_{ab}\lambda\nonumber\\
                                     &=& -\frac{i}{4}[(\partial_{\mu}\omega_{\nu}^{ab}-\partial_{\nu}\omega_{\mu}^{ab}) + (\omega_{\mu}^{ac}\omega_{\nu c}^{\quad b}-\omega_{\nu}^{ac}\omega_{\mu c}^{\quad b})]\sigma_{ab}\lambda.
\end{eqnarray}
Therefore, from equation \eqref{GCMDOA3}, it is possible to describe the commutation between the covariant derivatives acting on  ${\lambda}$ as:
\begin{equation}\label{GCMDOA4}
 [\nabla_{\mu}, \nabla_{\nu}]\lambda = -\frac{i}{4}F_{\mu\nu}^{~~~ab}\sigma_{ab}\lambda,
\end{equation}
where ${F_{\mu\nu}^{~~~ab} = \partial_{\mu}\omega_{\nu}^{ab}-\partial_{\nu}\omega_{\mu}^{ab} + \omega_{\mu}^{ac}\omega_{\nu c}^{\quad b}-\omega_{\nu}^{ac}\omega_{\mu c}^{\quad b}}$.

\section{From Einstein-Hilbert action to Palatini-Hilbert action with the term of Holst}\label{B}
\indent
\indent It is known, from the General Relativity, that the Riemann tensor can be obtained using the commutation of two covariant derivatives acting
on a vector, it means that
\begin{eqnarray}\label{GCMDOB1}
 [\nabla_{\mu},\nabla_{\nu}]V_{\rho} = R_{\nu ~ \mu\rho}^{~\alpha}V_{\alpha}\quad\textnormal{and}\quad R_{\nu ~ \mu\rho}^{~\alpha} = \partial_{\nu}\Gamma^{\alpha}_{\mu\rho} - \partial_{\mu}\Gamma^{\alpha}_{\nu\rho} + \Gamma^{\sigma}_{\rho\mu}\Gamma^{\alpha}_{\nu\sigma} - \Gamma^{\sigma}_{\rho\nu}\Gamma^{\alpha}_{\mu\sigma}.
\end{eqnarray}
The relation between the curvature tensor ${R_{\nu ~ \mu\rho}^{~\alpha}}$ and ${F_{\mu\nu}^{~~~ab}}$ originates from the commutation between the covariant derivatives under an internal index object (tetrad), ${[\nabla_{\mu},\nabla_{\nu}]S_{a}}$:
\begin{eqnarray}\label{GCMDOB2}
 [\nabla_{\mu},\nabla_{\nu}]S_{a} &=& \nabla_{\mu}\nabla_{\nu}S_{a} - \nabla_{\nu}\nabla_{\mu}S_{a} = [\partial_{\mu}(\partial_{\nu}S_{a}+\omega_{\nu a}^{\quad b}S_{b}) - \Gamma_{\mu\nu}^{\alpha}(\partial_{\alpha}S_{a}+\omega_{\alpha a}^{\quad b}S_{b})
 + \omega_{\mu a}^{\quad c}(\partial_{\nu}S_{c}+\omega_{\nu c}^{\quad b}S_{b})]\nonumber\\ &&-[\partial_{\nu}(\partial_{\mu}S_{a}+\omega_{\mu a}^{\quad b}S_{b}) - \Gamma_{\nu\mu}^{\alpha}(\partial_{\alpha}S_{a}+\omega_{\alpha a}^{\quad b}S_{b})+ \omega_{\nu a}^{\quad c}(\partial_{\mu}S_{c}+\omega_{\mu c}^{\quad b}S_{b})],
\end{eqnarray}
again, because of the torsion free scenario ${\Gamma_{\mu\nu}^{\alpha} - \Gamma_{\nu\mu}^{\alpha} = 0}$, we get
\begin{eqnarray}\label{GCMDOB3} 
 [\nabla_{\mu},\nabla_{\nu}]S_{a} &=& \partial_{\mu}\partial_{\nu}S_{a} + (\partial_{\mu}\omega_{\nu a}^{\quad b})S_{b} + \omega_{\nu a}^{\quad b}(\partial_{\mu}S_{b}) + \omega_{\mu a}^{\quad c}(\partial_{\nu} S_{c}) + \omega_{\mu a}^{\quad c}\omega_{\nu c}^{\quad b}S_{b}\nonumber\\
                                  &&-\partial_{\nu}\partial_{\mu}S_{a} - (\partial_{\nu}\omega_{\mu a}^{\quad b})S_{b} - \omega_{\mu a}^{\quad b}(\partial_{\nu}S_{b}) - \omega_{\nu a}^{\quad c}(\partial_{\mu} S_{c}) - \omega_{\nu a}^{\quad c}\omega_{\mu c}^{\quad b}S_{b}\nonumber\\
                                  &=&(\partial_{\mu}\omega_{\nu a}^{\quad b})S_{b} - (\partial_{\nu}\omega_{\mu a}^{\quad b})S_{b} + \omega_{\mu a}^{\quad c}\omega_{\nu c}^{\quad b}S_{b} - \omega_{\nu a}^{\quad c}\omega_{\mu c}^{\quad b}S_{b},
\end{eqnarray}
and it allows us to concluded that 
\begin{equation}\label{GCMDOB4}
 [\nabla_{\mu},\nabla_{\nu}]S^{a} = F_{\mu\nu}^{\quad ab}S_{b},
\end{equation}
where ${F_{\mu\nu}^{\quad ab} = \partial_{\mu}\omega^{ab}_{\nu} - \partial_{\nu}\omega^{ab}_{\mu} + \omega_{\mu}^{ac}\omega^{\quad b}_{\nu c} - \omega_{\nu}^{ac}\omega^{\quad b}_{\mu c}}$.\\
\indent Now, with the vector ${V_{\alpha}}$ in terms of tetrads and using the equation \eqref{GCMDOB4}, the expression \eqref{GCMDOB1} can be rewriting as
\begin{eqnarray}\label{GCMDOB5}
 R_{\nu ~ \mu\rho}^{~\alpha}V_{\alpha} = [\nabla_{\mu},\nabla_{\nu}]V_{\rho} = [\nabla_{\mu},\nabla_{\nu}]e_{\rho}^{a}V_{a} = e_{\rho}^{a}[\nabla_{\mu},\nabla_{\nu}]V_{a} = e_{\rho}^{a}F_{\mu\nu a}^{\quad b}V_{b} = e_{\rho}^{a}F_{\mu\nu a}^{\quad b}e^{\alpha}_{b}V_{\alpha},
\end{eqnarray}
then the relation between tensor curvature and ${F_{\mu\nu}^{\quad ab}}$ takes the form:
\begin{equation}\label{GCMDOB6}
 R_{\nu ~ \mu\rho}^{~\alpha} = e_{\rho}^{a}F_{\mu\nu a}^{\quad b}e^{\alpha}_{b}.
\end{equation}
From equation \eqref{GCMDOB6}:
\begin{eqnarray}\label{GCMDOB7}
 R_{\nu\sigma\mu\rho} &=& g_{\sigma\alpha}R_{\nu~ \mu\rho}^{~\alpha} = g_{\sigma\alpha}e_{\rho}^{a}F_{\mu\nu a}^{\quad b}e^{\alpha}_{b} = \eta_{cd}e_{\sigma}^{c}e^{d}_{\alpha}e_{\rho}^{a}F_{\mu\nu a}^{\quad b}e^{\alpha}_{b} = \eta_{cd}e_{\sigma}^{c}e_{\rho}^{a}F_{\mu\nu a}^{\quad b}\delta_{b}^{d} = \eta_{cd}e_{\sigma}^{c}e_{\rho}^{a}F_{\mu\nu a}^{\quad d}= R_{\mu\rho\nu\sigma}.
\end{eqnarray}
Once the Ricci's scalar is ${R = g^{\mu\nu}R_{\mu\nu} = R_{\mu\rho\nu\sigma}g^{\mu\nu}g^{\rho\sigma}}$, from the equation \eqref{GCMDOB7} it follows that
\begin{eqnarray}\label{GCMDOB8}
 R &=& \eta_{cd}e_{\sigma}^{c}e_{\rho}^{a}F_{\mu\nu a}^{\quad d}\eta^{gh}e_{g}^{\mu}e_{h}^{\nu}\eta^{jk}e_{j}^{\rho}e_{k}^{\sigma} = \eta_{cd}e_{\sigma}^{c}e_{\rho}^{a}F_{\mu\nu a}^{\quad d}\eta^{ch}e_{c}^{\mu}e_{h}^{\nu}\eta^{ak}e_{a}^{\rho}e_{k}^{\sigma}\nonumber\\
   &=& \eta_{cd}\eta^{ch}\eta^{ak}e_{\sigma}^{c}e_{c}^{\mu}e_{\rho}^{a}e_{a}^{\rho}e_{h}^{\nu}e_{k}^{\sigma}F_{\mu\nu a}^{\quad d} = \delta_{d}^{h}\delta_{\sigma}^{\mu}e_{h}^{\nu}e_{k}^{\sigma}\eta^{ak}F_{\mu\nu a}^{\quad d}\nonumber\\
   &=&e^{\nu}_{d}e^{\mu}_{k}\eta^{ak}F_{\mu\nu a}^{\quad d}\nonumber\\
   &=&e^{\nu}_{d}e^{\mu}_{k}F_{\mu\nu}^{\quad kd}.
\end{eqnarray}
Making the change ${k\rightarrow a}$ and ${d\rightarrow b}$, the scalar of curvature, via tetrads and connection ${(e, \omega)}$, is
\begin{equation}\label{GCMDOB9}
 R = F_{\mu\nu}^{~~~ ab}e_{a}^{\mu}e_{b}^{\nu}.
\end{equation}
\indent The Jacobian term for volume correction in curved spaces, ${\sqrt{-g}}$, in terms of tetrads is characterized by
\begin{eqnarray}\label{GCMDOB10}
 \sqrt{-g} = [-det(\eta_{ab}e_{\mu}^{a}e_{\nu}^{b})]^{1/2} = [-det(\eta_{ab})det(e_{\mu}^{a})det(e_{\nu}^{b})]^{1/2} = [det(e_{\mu}^{a})det(e_{\nu}^{b})]^{1/2} = (e^{2})^{1/2} = e.
\end{eqnarray}
 Thus, Einstein-Hilbert action can be written with the Riemann-Cartan formalism, via equations \eqref{GCMDOB9} and \eqref{GCMDOB10}, in the so-called Palatini-Hilbert action \cite{Palatini}:
\begin{equation}\label{GCMDOB11}
 \mathcal{S}_{EH} = \frac{1}{16\pi G}\int_{\mathcal{M}}d^{4}x\sqrt{-g}R\longmapsto\mathcal{S}_{PH} = \frac{1}{16\pi G}\int_{\mathcal{M}}d^{4}x(e)F_{\mu\nu}^{~~~ ab}e_{a}^{\mu}e_{b}^{\nu}.
\end{equation}
Almost 77 years after the Palatini's work, Holst obtained a dual term associated with curvature ${F_{\mu\nu}^{\quad ab}}$ in the
Hamiltonian formalism for gravity, now called the term of Holst (self-dual of Palatini action) \cite{Holst:1995pc}:
\begin{equation}\label{GCMDOB12}
 -\frac{1}{2\gamma}ee_{I}^{\mu}e_{J}^{\nu}F_{\mu\nu}^{~~~ IJ}(\omega),
\end{equation}
where ${(I,J)}$ are the spatial components of Lorentz internal indices ${(a,b)}$ and ${\gamma}$ is the Barbero-Immirzi parameter, 
proposed by both authors in refs. \cite{Barbero:1994ap, Immirzi:1996di}, in which the self-dual formulation corresponds to the choice of ${\gamma = -i}$.
This Holst term allows the Palatini-Hilbert action to be generalized to
\begin{equation}\label{GCMDOB13}
 \mathcal{S}_{PH} = \frac{1}{16\pi G}\int_{\mathcal{M}}d^{4}x(ee^{\mu}_{I}e^{\nu}_{J}P^{IJ}_{~~~ KL}F_{\mu\nu}^{~~~KL}(\omega)),
\end{equation}
being that, this value ${\gamma = \pm i}$ can not be performed to deal with its inverse:

\begin{eqnarray}\label{GCMDOB14}
 P^{IJ}_{~~~KL} = \delta_{K}^{[I}\delta_{L}^{J]} - \frac{1}{\gamma}\frac{\epsilon^{IJ}_{~~~KL}}{2}\quad\textnormal{and}\quad P^{-1~KL}_{IJ} = \frac{\gamma^{2}}{\gamma^{2}+1}\left(\delta_{I}^{[K}\delta_{J}^{L]} + \frac{1}{\gamma}\frac{\epsilon^{~~~ KL}_{IJ}}{2}\right). 
\end{eqnarray}

\section{Palatini-Hilbert-Holst coupling MDO action in ADM and tetrad formalism}\label{C}
\indent
\indent Knowing that ${|det(e_{a}^{i})| = \sqrt{q}}$ and ${|e| = |\sqrt{-g}| = N\sqrt{q}}$, and using the decomposition \eqref{GCMDO15}, 
the gravitational Lagrangian density of action \eqref{GCMDO11} is expressed as ${ee^{\mu}_{I}e^{\nu}_{J}P^{IJ}_{~~~ KL}F_{\mu\nu}^{~~~ KL}(\omega) \mapsto N\sqrt{q}e^{a}_{I}e^{b}_{J}P^{IJ}_{~~~ KL}F_{ab}^{~~~ KL}(\omega)}$:
\begin{eqnarray}\label{GCMDOC1}
 N\sqrt{q}e^{a}_{I}e^{b}_{J}P^{IJ}_{~~~KL}F_{ab}^{~~~KL}(\omega) &=& N\sqrt{q}(\varepsilon_{I}^{a} - n^{a}n_{I})(\varepsilon_{J}^{b} - n^{b}n_{J})P^{IJ}_{~~~KL}F_{ab}^{~~~KL}(\omega)\nonumber\\
                              &=& N\sqrt{q}(\varepsilon^{a}_{I}\varepsilon^{b}_{J} - \varepsilon_{I}^{a}n^{b}n_{J} - n^{a}n_{I}\varepsilon_{J}^{b} + n^{a}n_{I}n^{b}n_{J})P^{IJ}_{~~~ KL}F_{ab}^{~~~KL}(\omega),
\end{eqnarray}
where the last three terms inside parentheses are, from equation \eqref{GCMDO16},
\begin{eqnarray}\label{GCMDOC2}
n^{a}n_{I}n^{b}n_{J} - \varepsilon_{I}^{a}n^{b}n_{J} - n^{a}n_{I}\varepsilon_{J}^{b} &=& (N^{-1}(t^{a} - N^{a}))n_{I}(N^{-1}(t^{b} - N^{b}))n_{J} - \varepsilon_{I}^{a}(N^{-1}(t^{b} - N^{b}))n_{J}\nonumber\\
                                                                                     &&- (N^{-1}(t^{a}- N^{a}))n_{I}\varepsilon_{J}^{b}\nonumber\\
                                                                                     &=& N^{-2}(t^{a}-N^{a})(t^{b}-N^{b})(\delta_{I}^{0}\delta_{J}^{0}) - N^{-1}\varepsilon_{I}^{a}t^{b}n_{J} + N^{-1}\varepsilon_{I}^{a}N^{b}n_{J}\nonumber\\
                                                                                     &&-N^{-1}t^{a}n_{I}\varepsilon_{J}^{b} + N^{-1}N^{a}n_{I}\varepsilon_{J}^{b}\nonumber\\
                                                                                     &=& -N^{-1}[\varepsilon_{I}^{a}t^{b}n_{J}+\varepsilon_{J}^{b}t^{a}n_{I}] + N^{-1}[N^{a}n_{I}\varepsilon_{J}^{b}+N^{b}n_{J}\varepsilon^{a}_{I}]\nonumber\\
                                                                                     &=& -2N^{-1}n_{I}t^{a}\varepsilon_{J}^{b} + 2N^{-1}N^{a}n_{I}\varepsilon_{J}^{b},
\end{eqnarray}
so that, using the equations \eqref{GCMDOC2} in \eqref{GCMDOC1}, the gravitational Lagrangian density becomes
\begin{equation}\label{GCMDOC3}
ee^{\mu}_{I}e^{\nu}_{J}P^{IJ}_{~~~KL}F_{\mu\nu}^{~~~KL}(\omega) \mapsto N\sqrt{q}(\varepsilon_{I}^{a}\varepsilon_{J}^{b}-2N^{-1}n_{I}t^{a}\varepsilon_{J}^{b} + 2N^{-1}N^{a}n_{I}\varepsilon_{J}^{b})P^{IJ}_{~~~KL}F_{ab}^{~~~KL}(\omega). 
\end{equation}
Following the same prescription, the parts with similar characteristic to scalar-S and quadratic fermionic-F dynamics, due to nature of MDO spinor, in action \eqref{GCMDO11}, are obtained by
\begin{equation}\label{GCMDOC4}
 e[(e^{\mu}_{I}e^{\nu}_{J}\eta^{IJ}\partial_{\mu}\stackrel{\neg}{\lambda}\partial_{\nu}\lambda) - m^{2}\stackrel{\neg}{\lambda}\lambda] \mapsto N\sqrt{q}[(\varepsilon_{I}^{a}\varepsilon_{J}^{b}-2N^{-1}n_{I}t^{a}\varepsilon_{J}^{b} + 2N^{-1}N^{a}n_{I}\varepsilon_{J}^{b})\eta^{IJ}(\partial_{a}\stackrel{\neg}{\lambda}\partial_{b}\lambda)-m^{2}\stackrel{\neg}{\lambda}\lambda] 
\end{equation}
and
\begin{eqnarray}\label{GCMDOC5}
 &&ee^{\mu}_{I}e^{\nu}_{J}\eta^{IJ}[\stackrel{\neg}{\lambda}\omega_{\mu}^{IJ}\sigma_{IJ}(\partial_{\nu}\lambda) - (\partial_{\mu}\stackrel{\neg}{\lambda})\omega_{\nu}^{MN}\sigma_{MN}\lambda - i/4\stackrel{\neg}{\lambda}(\omega_{\mu}^{IJ}\omega_{\nu}^{MN}\sigma_{IJ}\sigma_{MN})\lambda]\mapsto\nonumber\\
 &&N\sqrt{q}[\varepsilon_{I}^{a}\varepsilon_{J}^{b}-2N^{-1}n_{I}t^{a}\varepsilon_{J}^{b} + 2N^{-1}N^{a}n_{I}\varepsilon_{J}^{b}]\eta^{IJ}\{i/2(\stackrel{\neg}{\lambda}\omega_{a}^{IJ}[\gamma_{I}\gamma_{J}-\gamma_{J}\gamma_{I}](\partial_{b}\lambda)- (\partial_{a}\stackrel{\neg}{\lambda})\omega_{b}^{MN}[\gamma_{M}\gamma_{N}-\gamma_{N}\gamma_{M}]\lambda)\nonumber\\
 &&+i/16\stackrel{\neg}{\lambda}(\omega_{a}^{IJ}\omega_{b}^{MN}[\gamma_{I}\gamma_{J}-\gamma_{J}\gamma_{I}][\gamma_{M}\gamma_{N}-\gamma_{N}\gamma_{M}])\lambda\}.
\end{eqnarray}
Thus, from the expressions \eqref{GCMDOC3}, \eqref{GCMDOC4} and \eqref{GCMDOC5}, the action of MDO matter with the Palatini-Holst gravitation \eqref{GCMDO11} is described in ADM formalism and tetrad field as
\begin{eqnarray}\label{GCMDC6}
 \mathcal{S}[\varepsilon, \omega, \lambda] &=& \mathcal{S}_{PH,\xi} + \mathcal{S}_{\lambda, \stackrel{\neg}{\lambda}_{S}} + \mathcal{S}_{\lambda, \stackrel{\neg}{\lambda}_{F}}\nonumber\\
                                 &=&\frac{1}{16\pi G}\int_{\mathbb{R}}dt\int_{\Sigma}d^{3}xN\sqrt{q}\Omega_{IJ}^{ab}P^{IJ}_{~~~KL}F_{ab}^{~~~KL}(\omega)(1-8\pi G\xi\stackrel{\neg}{\lambda}\lambda)\nonumber\\
                                 &&+\frac{1}{2}\int_{\mathbb{R}}dt\int_{\Sigma}d^{3}xN\sqrt{q}[\Omega_{IJ}^{ab}\eta^{IJ}(\partial_{a}\stackrel{\neg}{\lambda}\partial_{b}\lambda)-m^{2}\stackrel{\neg}{\lambda}\lambda]\nonumber\\
                                 &&-\frac{1}{8}\int_{\mathbb{R}}dt\int_{\Sigma}d^{3}xN\sqrt{q}\Omega_{IJ}^{ab}\eta^{IJ}\{i/2(\stackrel{\neg}{\lambda}\omega_{a}^{IJ}[\gamma_{I},\gamma_{J}](\partial_{b}\lambda)\nonumber\\
                                 &&- (\partial_{a}\stackrel{\neg}{\lambda})\omega_{b}^{MN}[\gamma_{M},\gamma_{N}]\lambda)+i/16\stackrel{\neg}{\lambda}(\omega_{a}^{IJ}\omega_{b}^{MN}[\gamma_{I},\gamma_{J}][\gamma_{M},\gamma_{N}])\lambda\},
\end{eqnarray}
where $\Omega_{IJ}^{ab} = (\varepsilon_{I}^{a}\varepsilon_{J}^{b}-2N^{-1}n_{I}t^{a}\varepsilon_{J}^{b} + 2N^{-1}N^{a}n_{I}\varepsilon_{J}^{b})$. Being $S_{\lambda, \stackrel{\neg}{\lambda}_{S}}$ and $S_{\lambda, \stackrel{\neg}{\lambda}_{F}}$ the scalar and fermionic parts of the action, respectively.

\end{document}